# Metal-Enhanced Near-Infrared Fluorescence by Micropatterned Gold Nanocages

*Andrea Camposeo,[1] Luana Persano,[1] Rita Manco,[1,2] Yi Wang,[3,4] Pompilio Del Carro,[1] Chao Zhang,[5] Zhi-Yuan Li,[5,6,*] Dario Pisignano,[1,2,**] Younan Xia[3,***]*

[1]Istituto Nanoscienze-CNR, Euromediterranean Center for Nanomaterial Modelling and Technology (ECMT), via Arnesano I-73100, Lecce, Italy

[2]Dipartimento di Matematica e Fisica "Ennio De Giorgi", Università del Salento, via Arnesano I-73100 Lecce, Italy

[3]The Wallace H. Coulter Department of Biomedical Engineering, Georgia Institute of Technology and Emory University, Atlanta, Georgia, 30332, United States

[4]Key Laboratory of Green Synthesis and Applications, College of Chemistry, Chongqing Normal University, Chongqing 401331, P. R. China

[5]Laboratory of Optical Physics, Institute of Physics, Chinese Academy of Sciences, Beijing 100190, P. R. China

[6]College of Physics and Optoelectronics, South China University of Technology, Guangzhou 510641, P. R. China

KEYWORDS: NANOCAGES, NEAR-INFRARED EMISSION, FLUORESCENCE ENHANCEMENT, SOFT LITHOGRAPHY

ABSTRACT. In metal-enhanced fluorescence (MEF), the localized surface plasmon resonances of metallic nanostructures amplify the absorption of excitation light and assist in radiating the consequent fluorescence of nearby molecules to the far-field. This effect is at the base of various technologies that have strong impact on fields such as optics, medical diagnostics and biotechnology. Among possible emission bands, those in the near-infrared (NIR) are particularly intriguing and widely used in proteomics and genomics due to its noninvasive character for biomolecules, living cells, and tissues, which greatly motivates the development of effective, and eventually multifunctional NIR-MEF platforms. Here we demonstrate NIR-MEF substrates based on Au nanocages micropatterned with a tight spatial control. The dependence of the fluorescence enhancement on the distance between the nanocage and the radiating dipoles is investigated experimentally and modeled by taking into account the local electric field enhancement and the modified radiation and absorption rates of the emitting molecules. At a distance around 80 nm, a maximum enhancement up to 2-7 times with respect to the emission from pristine dyes (in the region 660 nm-740 nm) is estimated for films and electrospun nanofibers. Due to their chemical stability, finely tunable plasmon resonances, and large light absorption cross sections, Au nanocages are ideal NIR-MEF agents. When these properties are integrated with the hollow interior and controllable surface porosity, it is feasible to develop a nanoscale system for targeted drug delivery with the diagnostic information encoded in the fluorophore.

Metal-enhanced fluorescence (MEF) is an optical process that has attracted increasing attention over the past decade, for its applications in many areas of optics, molecular physics, medical diagnostics and biological research.[1-7] For MEF to occur, a fluorophore has to be placed in close proximity to a metal surface. The localized surface plasmon resonance (LSPR) of the metal can strongly amplify the absorption of incident light used for photo-excitation, and the energy can be transferred to the fluorophore if the LSPR band has a substantial overlap with the excitation spectrum of the molecule.[4] Metal nanoparticles are particularly intriguing MEF antennas, since they can be used to generate significant local enhancement of the surrounding electromagnetic field, especially at the tips and corners.[8,9] In addition, light scattering from the metal can assist in radiating the fluorescence emission from the fluorophore to the far-field. As a result of these mechanisms, the brightness and quantum yield of the fluorophore increase, as well as its photostability.[3,7] In the past, various types of metal nanoparticles and surfaces have been examined as MEF agents. These include systems made of Au,[10-14] Ag,[8,9,15-18] and Cu,[19] with a variety of shapes such as spheres,[11,12,19,18] rods,[9,14] plates,[8] prisms,[15] shells,[13,14] and cubes,[16] as well as nanoporous and continuous films.[10,17] Most recently, Ag nanocubes deposited on metal films with a nano-sized gap loaded with emitters has been demonstrated as an excellent nanoscale patch antenna for large Purcell enhancement.[20]

Among the different emission bands of available fluorophores, near-infrared (NIR, with wavelengths in the range 650-900 nm) is especially important because it is commonly used in proteomic and genomic analysis, it is noninvasive with respect to living tissues, and it generally shows reduced autofluorescence and deep penetration for soft tissues.[21,22] This spectral region is also known as "water window" due to the minimal light absorption from water (absorbing for wavelengths above 900 nm) and, at the same time, from hemoglobin (absorbing for wavelengths below 650 nm), thus offering high transparency when imaging soft tissues and blood.[23] The distinct features of NIR over UV and visible fluorescence also include the generally lower background signal related to autofluorescence from organic materials, as well as a largely reduced possibility of photo-induced damage. These features are remarkable when dealing with molecular samples and chromophores. On the other hand, achieving bright emission with NIR fluorescent molecules is generally much more difficult than with UV or visible chromophores and dyes. Hence, effective MEF-platforms allowing NIR fluorescence to be enhanced are very desirable, and can open up new routes for high-performance diagnostic devices and photonics.

Most of the MEF studies have been hitherto focused on fluorophores emitting in the visible region. Among the very few attempts targeting NIR, Anderson *et al*. studied Ag films deposited from $AgNO_3$ solutions and coated with dye-labeled DNA oligonucleotides and proteins.[24] Bardhan and colleagues studied Au nanoshells on silica particles and Au nanorods, to which indocyanine green or other chromophores were conjugated by involving silica or serum albumin spacers.[13,14] In designing such experiments, one has to address a number of issues, including engineering of the LSPR frequency and experimental achievement of the nanoscale distance between the metal and the fluorophore required for MEF. For these reasons, developed processes generally involve multiple and complex steps for particle functionalization and conjugation. In addition, the MEF agents used often lack the potential for multifunctionality, namely for inducing bright NIR-fluorescence and eventually carrying other diagnostic or therapeutic molecules simultaneously.

In this work we address these issues by demonstrating MEF in the NIR region based on Au nanocages, which are straightforwardly incorporated into a well-defined, microprinted structure with tight spatial control. Due to their chemical stability, finely tunable LSPR in the NIR, and large light absorption cross-sections (much higher than typical organic dyes),[25] Au nanocages are ideal MEF agents. The LSPR peaks of Au nanocages can be tuned from the visible to the NIR region by simply controlling their size and wall thickness.[25] Surprisingly, these properties have never been exploited before to develop nanocages as effective MEF agents. Here MEF was demonstrated in NIR light-emitting films and electrospun polymer nanofibers, with additional spatial control provided by microprinting of nanocages. The dependence of the enhancement on the distance from the cage to the radiating dipoles can be predicted based on the local electric field enhancement and the modified radiation and absorption rates of the emitting molecules. The combination with their hollow interior and controllable surface porosity for carrying drugs for targeted delivery, as well as their potential use as photothermal transducers,[26,27] makes nanocages a unique, multifunctional platform for diagnostics, therapeutics, and photonics.

### Results and Discussion

Figure 1a shows a schematics of the procedure used for fabricating the MEF substrates, which consist of patterned Au nanocages, covered by an e-beam deposited $SiO_2$ spacing layer, and by a dye-doped light-emitting polymer film or nanofiber. The Au nanocages with a hollow and porous structure were prepared in an aqueous solution through a galvanic replacement reaction that involved the use of Ag nanocubes as a sacrificial template and $HAuCl_4$ as a precursor. Figure 1b shows a transmission electron microscopy (TEM) image of the as-obtained Au nanocages. They have an average edge length of $(70.2 \pm 4.6)$ nm. The silica layer has good optical transparency, low surface roughness of ~ 1 nm, refractive index of about 1.45.[28] It keeps the fluorophores out of direct contact with the Au surface, where quenching might dominate over fluorescence enhancement. The LD 700 perchlorate dye is chosen as the fluorescent probe because it is one of the most commonly used fluorophores for DNA staining, flow cytometry, as well as low-threshold, mode-locked, and micro-lasers.[29,30] In addition, its spectral characteristics match the response range of photomultiplier detectors and arrays typically used for confocal imaging and visible-NIR spectroscopy (see Methods). The dye absorption and emission spectra are shown in Figure 1c, together with the broad LSPR band of the Au nanocages. If needed, the LSPR peaks of the resultant Au nanocages could be precisely tuned to match the fluorescence emission peaks of most dyes by titrating the Ag nanocubes with different amounts of an aqueous $HAuCl_4$ solution or using Ag nanocubes with different sizes as the templates.[31,32]

*Micropatterning and spacing control.* The nanocages were immobilized on quartz substrates as patterned arrays with micrometer resolution. To this end, the surface of a quartz substrate was patterned with (3-aminopropyl)triethoxysilane (APTES) through microcontact printing. The silane molecules were patterned as an array of squares 50 μm in edge length, followed by the selective immobilization of Au nanocages onto the APTES regions. The nanocages are then stably bonded providing a uniform surface coverage of (20±2)% (see Fig. 2a,b and Fig. S1 in the Supporting Information). The degree of surface coverage by nanoparticles mainly depends on the effectiveness of APTES microcontact printing, and on the concentration of nanocages used for immobilization. To create such uniform patterns, we have to pre-treat the substrates with piranha solution and oxygen plasma to favor the formation of stable bonds between the silanes and quartz, and sonicate the suspension of nanocages to avoid clustering or agglomeration. Once the MEF-substrates based on nanocages have been prepared and then coated with $SiO_2$ layers, various polymer films and nanostructures can be deposited on top of them. The substrates can be utilized many times, with different types of organic patterns. To highlight the feasibility of combining nanocage-based MEF substrates with different nanostructures, we also exploit NIR-light emitting polymer nanofibers directly deposited on the silica layer. The nanofibers are fabricated by electrospinning polyacrylonitrile (PAN) doped with LD 700, whose emission can be precisely modulated by the pattern of nanocages underneath (Fig. S2 in the Supporting Information).

It is well known that MEF critically depends on the distance between the metal surface and the fluorophores.[1-3] To investigate this issue in depth in our system, we systematically varied the thickness of the deposited silica layer, to move through the different regimes of interaction between the nanocages and NIR fluorophores. The fluorophores were deposited from a chloroform solution of poly(methyl methacrylate) (PMMA) containing LD 700, by spin-casting onto the SiO$_2$ layer. The resulting samples with different nanocage-LD 700 distances were then observed by laser confocal microscopy and the images are shown in Figure S3. For distances below 50 nm, fluorescence quenching is clearly observed, whereas for distance beyond 50 nm MEF is found with a maximum enhancement at a nanocage-fluorophore spacing of 80 nm. A maximum enhancement up to about 7 times with respect to the emission from pristine dyes is estimated, by taking into account the coverage of nanoparticles on the substrate. The dependence of the enhancement factor on the wavelength was also studied for different values of the spacing distance (50, 80 and 110 nm), that is, in the different regimes of nanocages-fluorophore interactions. Generally, maximum MEF can be seen at wavelengths in the region of 660-740 nm (Fig. 3a). The corresponding LD 700 emission spectra under pristine conditions, and under MEF at an emitter-nanocage separation of 80 nm are shown in Figure 3b.

*Nanocage-enabled MEF.* In order to rationalize the observed behavior of nanocage-enabled MEF for LD 700, one has to consider that the fluorescence rate, $\Gamma_{fl}^0$, of an emitter is determined by its own excitation rate, $\Gamma_{exc}^0$, and the intrinsic quantum yield, $\eta^0$, which is defined by $\eta^0 = \Gamma_{rad}^0 / (\Gamma_{rad}^0 + \Gamma_{nr}^0)$,[33] where $\Gamma_{rad}^0$ and $\Gamma_{nr}^0$ are the intrinsic radiative and nonradiative relaxation rate of the emitter in free space, respectively ($\Gamma_{fl}^0 = \Gamma_{exc}^0 \cdot \eta^0$). In particular, the fluorescence rate will be modified when the emitter is positioned next to a nanostructure, due to the scattering from the nanostructure. The modification can take place through two physical processes: fluorescence excitation and radiation. The excitation process is related to the transition rate from the ground state to the excited state of the dye. Such a rate is square-proportional to the transition matrix element $|\mathbf{p} \cdot \mathbf{E}|$, with $\mathbf{p}$ and $\mathbf{E}$ being the molecular dipole moment and local electric field, respectively. Hence, the excitation enhancement can be simply linked to the enhancement in local field intensity. Indeed, from the quantum theory of fluorescence transition, the excitation modification factor can be written as $\Gamma_{exc} / \Gamma_{exc}^0 = |E|^2 / |E_0|^2$, where $E_0$ is the incident field. The excitation modification factor is thus directly related to the local field enhancement factor, which can be readily calculated by means of electromagnetic simulations using, for instance, the finite-difference time-domain (FDTD) method. In our calculation, the surface on which nanocages are deposited is defined as the plane for the $x$ and $y$ axes, whereas its normal direction is labeled as $z$, as schematically illustrated in Figure 4a. A $z$-axis incident plane wave with $y$-polarization electric field is used to excite the metallic nanostructure, from which the local field as well as the fluorescence excitation modification factor at the wavelength, $\lambda_{exc}$ = 640 nm, can be obtained. Then we use a dipole source emitting at the maximum dye emission wavelength ($\sim$ 675 nm) and placed next to the nanostructure, to calculate the local field, from which the modified radiation and absorption rates of the molecules ($\Gamma_{rad} / \Gamma_{rad}^0$ and $\Gamma_{abs} / \Gamma_{rad}^0$), are derived to establish the modified quantum yield $(\eta/\eta_0)$. The modified factor for the fluorescence rate of molecules with a given orientation is then written as:

$$g = \frac{\Gamma_{fl}}{\Gamma_{fl}^0} = \frac{\Gamma_{exc}}{\Gamma_{exc}^0} \cdot \frac{\eta}{\eta_0} = \frac{\Gamma_{exc}}{\Gamma_{exc}^0} \cdot \frac{\Gamma_{rad}/\Gamma_{rad}^0}{\eta_0(\Gamma_{rad}/\Gamma_{rad}^0 + \Gamma_{abs}/\Gamma_{rad}^0) + (1-\eta_0)}. \tag{1}$$

The ultimate enhancement factor ($G$) is also affected by the collection angle associated with the experimental set-up ($\cong 24°$ in our experiments), which critically depends on the particular orientation of the emitting dipole in space:

$$G = \left(\sum_{i=x,y,z} \alpha_i g_i \cdot C_i\right) \bigg/ \sum_{i=x,y,z} \alpha_i C_i^0. \tag{2}$$

In this expression, we sum on the three spatial directions and introduce the collection efficiency coefficients with (without) the nanostructure, $C$ ($C^0$), and the coefficients, $\alpha_i$, which indicate the amount of emitting dipoles with orientation along each one of the three axes (the subscript, $i$, indicates the orientation of the dye molecule).

Firstly, we performed simulations for nanocages with different edge lengths in the range of 50-80 nm and different wall thicknesses in the range of 4-8 nm (Fig. S4). The nanocage was surrounded by a dielectric medium made of SiO$_2$, and our goal was to find the most suitable geometric parameters for the nanocage that would give an extinction spectrum analogous to that measured experimentally. The edge length derived from such a study was 70 nm, together with a wall thickness of 6 nm (Fig. S5). This allowed us to calculate the local field enhancement, $\Gamma_{exc}/\Gamma_{exc}^0$, induced by the nanocage (Fig. 4b). This enhancement is found to rapidly attenuate when one moves away from the particle (Fig. S6). We then calculated the modified radiation and absorption rates of the molecules with different spatial orientations and at different distances from the nanocage ($l$=15, 35, 55, 75, 95 and 115 nm). The results are shown in Figures S7 and S8, respectively. The ultimate enhancement factor for the dye molecules is determined, whose spatial variation is found to critically depend on the pristine quantum yield of the emitter. For a low-efficiency dye ($\eta_0 \cong 0$), the resulting curve does not fit the experimental one regardless of the light polarization considered (Figure S9), whereas for dyes with higher intrinsic quantum yields one can get curves which more closely resemble those from experimental measurements. In general, the fluorescence radiation enhancement is expected to be especially sensitive to the pristine efficiency of the dye used as discussed above. We find that for $x$-polarized emitting dipoles a lower efficiency leads to a higher enhancement regardless of the nanocage-emitter distance (Fig. 5a), whereas for $z$-polarized dipoles an increasing or decreasing trend is obtained for the enhancement upon decreasing $\eta_0$ for values of $l$ below or above 50 nm, respectively (Fig. 5b).

We also consider variations to the final enhancement factor upon varying the coordinate, $x_{NC}$, of the nanocage, namely dislocating the particle at a distance up to 0.5 μm along the $x$-direction (while keeping the emitting dipole at $x$=0). No dipole-nanocage interaction is effective for $x_{NC}$ larger than a few tens of nm (Fig. S10), suggesting that dye molecules in the PMMA films mainly interact with a single nanocage in our experimental configuration. This determines the in-plane resolution limit for the light-emitting patterns recorded from the nanocage-based MEF substrates. Finally, by taking into account the experimental collection angle as in Eq. (2), one could explain the experimental trends of nanocage-enabled MEF. Results from these simulations, in the prototypical cases of emitting dipoles lying along $z$ and of a population of emitting dipoles equally distributed along the $x$, $y$, and $z$ directions, are displayed in Figure 6. A quantitative agreement is obtained for nanocage-emitter distances below 60 nm, whereas for 60 nm < $l$ < 120 nm the experimental values of the fluorescence enhancement slightly outperform the ones from simulations (maximum calculated $G$ values are around 1.7 and 1.4 for $z$-polarized dipoles and for an isotropic distribution, respectively). In particular, the isotropic distribution of chromophore orientations in space, as expected for spin-coating deposition, leads to a dependence of $G$ on the nanocage-emitter distance which is in agreement with experimental data, with a well-predicted trend including quenching at small $l$ values and a maximum at $l \cong$ 60-80 nm followed by a decrease in $G$ at larger distances. Discrepancies in the obtained curves could be ascribed to a number of factors, including additional waveguiding and reflection effects of radiation at the involved optical interfaces. In particular, the deposition of the dielectric layer above the particles might lead to chromophore-nanocage distances spanning over a given range of separations. These would likely include separations below the nominal $l$ value, especially nearby the top corners of the nanocages, due to the shadow effect and local formation of silica grains during oxide evaporation, which would explain the observed deviation of the $G$ maximum with respect to theoretical calculations.

Finally, the extinction band position is found to affect significantly the NIR-MEF behavior, as evidenced by calculating the $l$-dependence of $G$ for nanocages of different size (e.g. 60 nm and 50 nm, with a wall thickness of 8 nm), therefore showing different spectral peaks (at 730 nm and 690 nm, respectively). In particular, we find that nanocages with a peak wavelength closer to that of the emitting dye lead to higher enhancement at relatively smaller distances, $l$. For instance, the maximum calculated $G$ values for an isotropic distribution of emitting dipoles and $l$ around 20-30 nm are ~1, 1.4, and 3.3, for nanocages with extinction band maxima at about 850, 730, and 690 nm, respectively (Fig. 6 and Figure S11). Overall, these results demonstrate that the microprinted nanocages are good MEF substrates with quite well predictable characteristics for utilization in biochips, diagnostics, and photonics.

**Conclusions**

We have systematically investigated the potential of Au nanocages as a novel class of MEF substrates in the NIR region. We deposited the Au nanocage as a patterned array on a solid substrate through microcontact printing. This allowed us to reliably examine the trend of the fluorescence enhancement factor as a function of the nanostructure-fluorophore distance. We also studied the enhancement in depth by numerical simulations and a reasonable agreement was achieved between the experimental and computational results. Applications of this technology will include planar immunoassays with nanoscale control over the diagnostic information and concomitant deep penetration capability into soft tissue specimens, development of new nanophotonic systems, and fabrication of nanoscale bar codes addressable by the nanocages and their patterns.

**Methods**

*Synthesis of Au nanocages*. The Au nanocages were prepared using a galvanic replacement reaction between Ag nanocubes (55 nm in edge length) and $HAuCl_4$ by following one of our previously reported protocols.[31] The reaction was monitored by measuring the LSPR peaks with a UV-vis spectrometer (Lambda 750, Perkin Elmer, Waltham, MA). The titration was terminated when the main LSPR peak of the Au nanocages had reached ~850 nm. The final products were collected by centrifugation and washing with saturated NaCl solution once and DI water three times. The geometric parameters of the Au nanocages such as the edge length (70.2 ± 4.6) nm and wall thickness (7.8 ± 1.5) nm were estimated using a transmission electron microscope (Hitachi 7500, Tokyo, Japan).

*Micropatterning of Au nanocages*. Masters with square patterns (period = 85 μm) were fabricated by photolithography with the use of $Si/SiO_2$ ($SiO_2$ thickness = 100 nm) substrates. After photoresist deposition (Clariant AZ5214E: AZ EBR Solvent 1:1), UV exposure (8' 30'' at 45 W) and development, we etched the thermal dioxide with $NH_4F/HF/H_2O$ (6.6 g/1.6 mL/10 mL). Poly(dimethysiloxane) (PDMS) stamps were then fabricated by pouring a mixture of Sylgard 184 elastomer and curing agent (w/w= 9:1) onto the masters, and heating at 140 °C for 15 min after degassing. To remove organic residues, the surface of quartz substrates were cleaned by immersion in a freshly prepared piranha solution (7:3, $H_2SO_4:H_2O_2$) for 60 min at 120 °C, and rinsed three times for 15 min in bidistilled water under ultrasonication at room temperature. Afterwards, the cleaned substrates were treated with an oxygen plasma at a power of 50 W for 1 min. A 10% APTES (Fluka) aqueous solution was deposited on the patterned side of a PDMS stamp for 30 min, which was then dried with a nitrogen stream and put in conformal contact with a cleaned quartz substrate for a few minutes. After peeling-off the PDMS stamp, the solution with Au nanocages was put in contact with the patterned surface for 5 min before rinsing several times with bidistilled water and nitrogen to remove unattached nanoparticles. Micropatterned nanocages were imaged by scanning electron microscopy (SEM, FEI Nova NanoSEM 450), and the degree of surface coverage by nanoparticles was evaluated by the WSxM software.[34]

The growth of $SiO_2$ layers on the substrates was then carried out by electron-beam evaporation using a Temescal Supersource 2 electronic gun system operated at 9 kV. The silica deposition was performed at 220 °C, with a rate of about 1.5 Å/s and a partial pressure of oxygen of $3\times10^{-5}$ mbar. The thickness of the deposited silica was monitored by a quartz microbalance and independently measured by both tapping-mode atomic force microscopy (AFM) and cross-sectional SEM at an acceleration voltage of 15 kV (Figure S12). AFM measurements were performed in air at 22°C, using a Nanoscope IIIa controller with a MultiMode head (Veeco). P-doped Si tips were used with 8 nm nominal curvature radius and resonant frequency of 190 kHz. Solutions of PMMA (Sigma-Aldrich)/LD 700 perchlorate (Exciton  Inc.) (w/w = 0.1%) in chloroform were mixed using an ultrasonicator for 2 h at 40°C, and by stirring for 12 h at room temperature, and then deposited by spin-coating.

*Electrospinning*. A solution of PAN/LD 700 perchlorate (w/w = 0.1%) in dimethylformamide (Sigma-Aldrich) ($\cong$ 65 mg/mL) was stirred overnight and then loaded into a syringe with a 23 gauge stainless steel needle. A 10 kV voltage is applied to the needle by a high-voltage power supply (EL60R0.6-22, Glassman High Voltage, High Bridge, NJ) and the injection rate was kept constant at 10 μL/min using a syringe pump (Harvard Apparatus, Holliston, MA). Electrospun fibers with diameters of 110 nm were deposited for a few seconds on the MEF substrates mounted on a grounded collector at a distance of 20 cm from the needle.

*Fluorescence measurements and analysis*. The absolute fluorescence quantum yield of the dye embedded in a PMMA matrix was measured using an integrating sphere, which involves excitation of the samples with a diode laser and collection of the emission with a fiber-coupled monochromator equipped with a charge-coupled device (CCD, USB 4000, Ocean Optics). This procedure provides an absolute quantum yield of 0.21±0.01 for the dyes embedded in the film. A 2-3fold higher quantum yield is generally observed for dyes dissolved in solutions relative to those in solid-state matrices.[35,36] Fluorescence micrographs of MEF patterns were obtained using an inverted microscope, Eclipse Ti, equipped with a confocal A1 R MP system (Nikon, Melville, NY). We excited the samples with a diode laser ($\lambda_{exc}$ = 640 nm) and detected the signals with a spectral detection unit equipped with a multi-anode photomultiplier with detection wavelength up to 750 nm (Nikon, Melville, NY). MEF and quenching measurements were performed for samples with adjacent Au nanocage-covered and pristine areas, in order to achieve the same conditions for substrate preparation and layer thickness and thus to ensure comparability of the data. Photoluminescence spectra were collected by exciting the samples with a diode laser ($\lambda_{exc}$=408 nm, beam diameter about 1 mm). The samples were mounted on a precision 3-axis translation stage, which allows specific regions of the samples to be selectively excited and characterized. The light emitted from the sample was collected along the direction normal to the active layer/air interface and coupled to a monochromator (iHR320, Jobin Yvon) equipped with a CCD (Symphony, Jobin Yvon) with overall spectral detection range up to 950 nm (as limited by the CCD quantum efficiency and grating diffraction efficiency). The measurements were performed in air and at ambient temperature.

*Electromagnetic simulation methods.* The fluorescence radiation of a fluorophore near the surface of a nanostructure is simulated by means of the FDTD method with three steps. In our FDTD simulations, the studied region was a 3 μm × 3 μm rectangle surrounded by perfectly matched layers, and the size of the mesh was set to 2 nm. First, we calculated the extinction spectrum and the local field enhancement of a nanocage excited by plane wave at an excitation wavelength 640 nm. Then we used an electric dipole to simulate the radiation of the dye molecule near the nanostructure. The electric dipole was set to 20-120 nm above the middle plane of the nanocage. By calculating the radiation and absorption power we obtained the total enhancement factor $g$ for a dye molecule at the radiation center wavelength of 675 nm. Finally, we obtained the collection efficiencies by calculating the energy flow through a monitoring area above the electric dipole with or without the nanostructure, and following Eq. 2 we obtained the curves shown in Figure 6.

ASSOCIATED CONTENT

**Supporting Information**.  A supporting document with additional details on SEM and confocal microscopy analysis of the microprinted Au nanocages, thickness measurements, and FDTD simulations is included as a separate PDF file. This material is available free of charge *via* the Internet at http://pubs.acs.org.

AUTHOR INFORMATION

**Corresponding Authors**

* lizy@iphy.ac.cn

** dario.pisignano@unisalento.it

*** younan.xia@bme.gatech.edu

## ACKNOWLEDGMENT

The research leading to these results has received funding from the European Research Council under the European Union's Seventh Framework Programme (FP/2007-2013)/ERC Grant Agreement n. 306357 (ERC Starting Grant "NANO-JETS"). The Apulia Networks of Public Research Laboratories Wafitech (09) and M. I. T. T. (13) are also acknowledged. Dr. Vito Fasano is acknowledged for absorption and photoluminescence measurements on pristine films. The samples of Au nanocages were prepared under financial support from the National Science Foundation (DMR 1215034) to Y. X..

## Figures and Captions

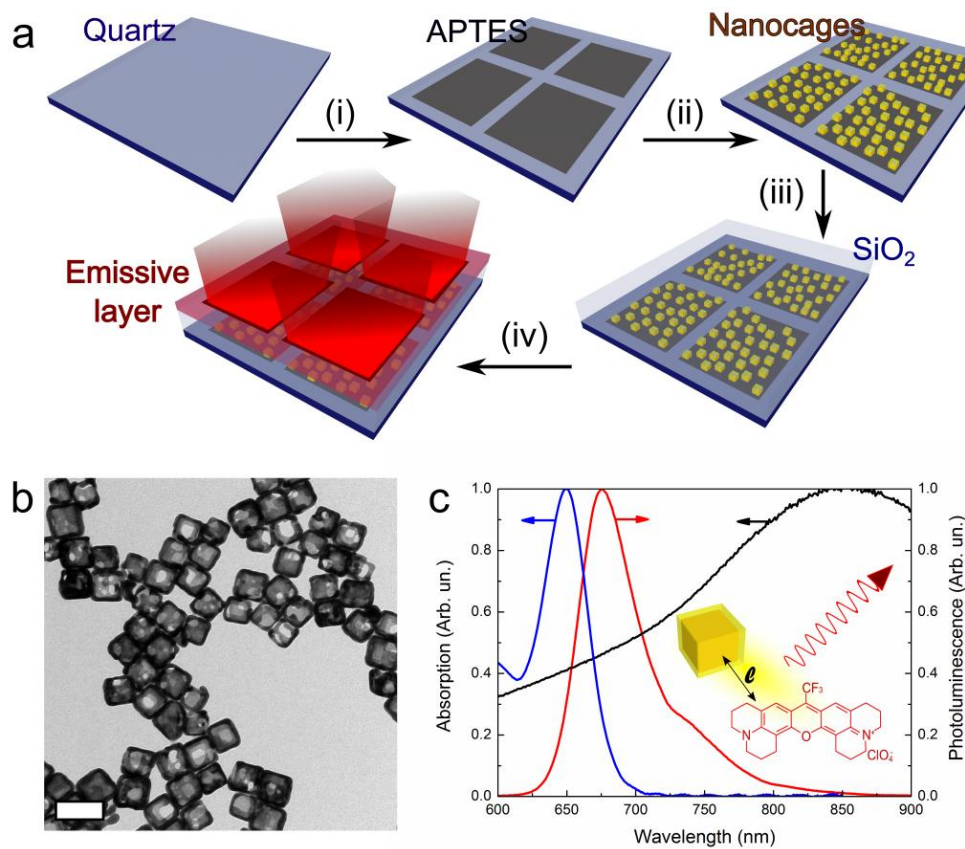

**Figure 1.** (a) A schematic showing all major steps involved in the fabrication of MEF substrates based on Au nanocages: (i) microcontact printing of APTES, (ii) immobilization of Au nanocages on the regions covered by APTES, (iii) deposition of a SiO₂ layer over the entire surface through e-beam evaporation, and (iv) deposition of an emissive layer to achieve a spatially-resolved MEF. (b) Transmission electron micrograph of the Au nanocages. Scale bar: 100 nm. (c) Absorption (blue curve) and emission (red curve) spectra of the LD 700 dye, and extinction spectrum (black curve) of the nanocages. Inset: schematic illustration of the MEF and molecular structure of LD 700. *l*: the distance between an emitter and the surface of the Au nanocage.

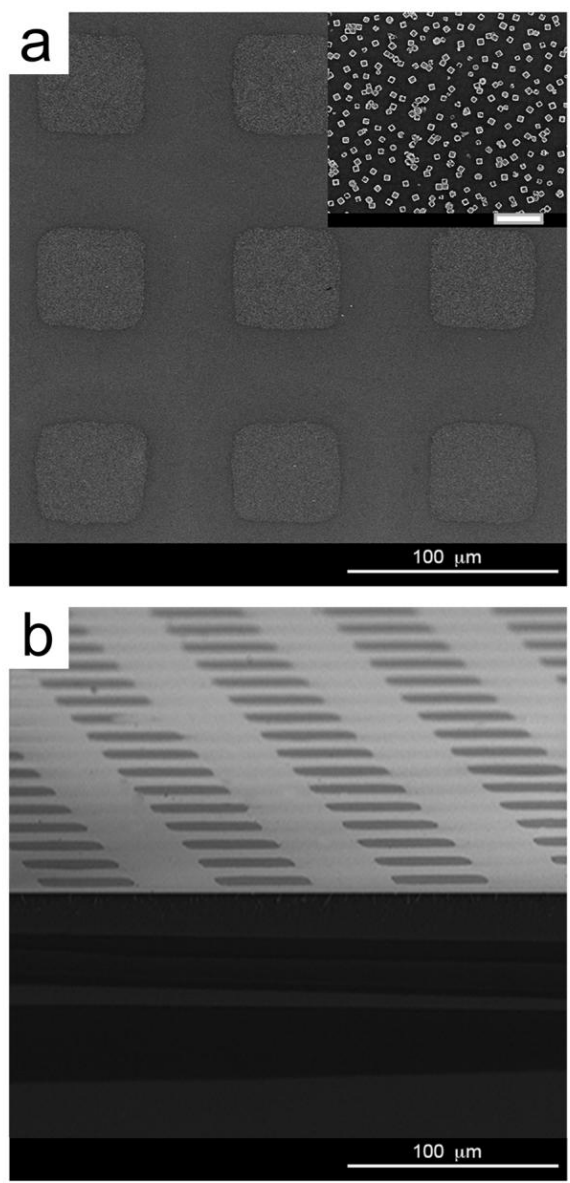

**Figure 2.** (a, b) Scanning electron micrograph of a patterned array of squared regions covered by Au nanocages, with each square having an edge length of 50 µm. The Au nanocages have an edge length of 70 nm. Inset in (a): image of an individual square at a higher magnification. Scale bar: 500 nm.

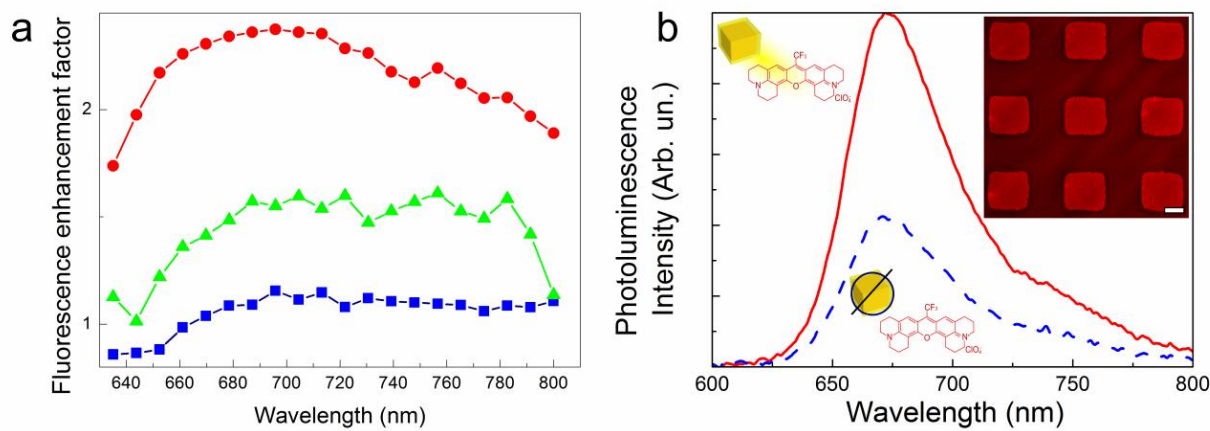

**Figure 3.** (a) Fluorescence enhancement factors measured for MEFs with $SiO_2$ layers of 50 nm (square dots), 80 nm (circular dots) and 110 nm (triangular dots), in thickness. (b) Corresponding photoluminescence spectra recorded from the dye molecules located in a region without (dashed line) and with Au nanocages (continuous line), highlighting the MEF-induced enhancement. The top-right inset shows an exemplary confocal micrograph of a patterned MEF substrate. Scale bar: 25 μm. A set of confocal images recorded from MEF substrates with different nanocage-emitter distances is shown in Fig. S3 of the Supporting Information.

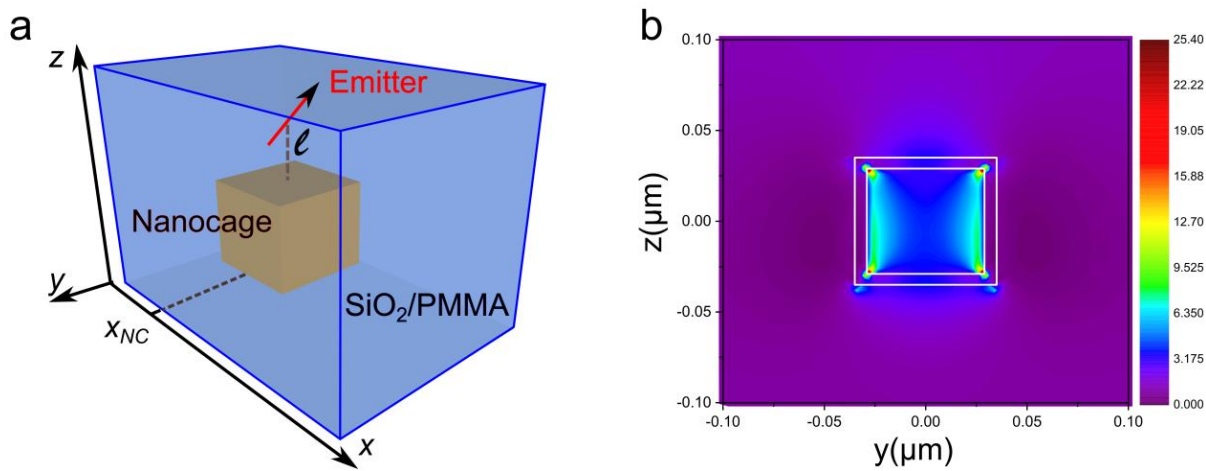

**Figure 4**. (a) A schematic of the MEF structure used in our simulations, in which a Au nanocage is embedded in a dielectric medium at $x = x_{CN}$. Here $l$ indicates the distance from the nanocage to an emitting dipole embedded in the dielectric medium. (b) Calculated local field enhancement around the nanocage.

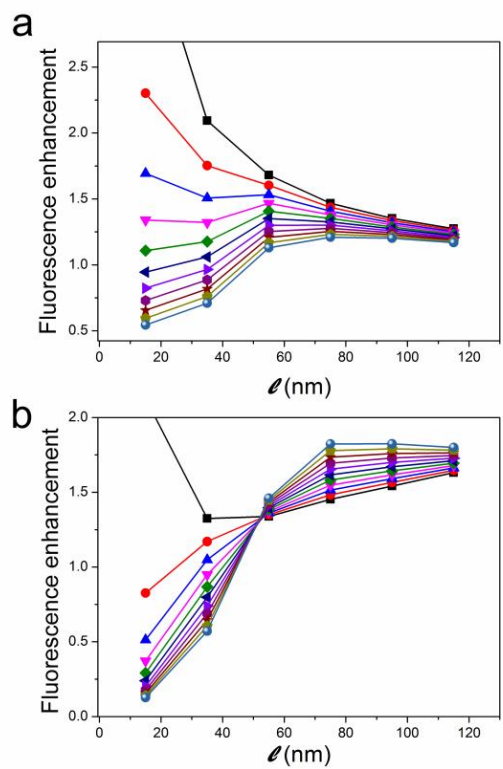

**Figure 5.** Calculated enhancement factor for a chromophore with (a) *x*-polarized and (b) *z*-polarized dipole and various nanocage-emitter distances and emitter yields. The nanocages have an edge length of 70 nm, and the collection angle is set to 24°. From top to bottom (referred to data points for *l*=15 nm): $\eta_0$ = 0, 0.1, 0.2, 0.3, 0.4, 0.5, 0.6, 0.7, 0.8, 0.9, 1.0.

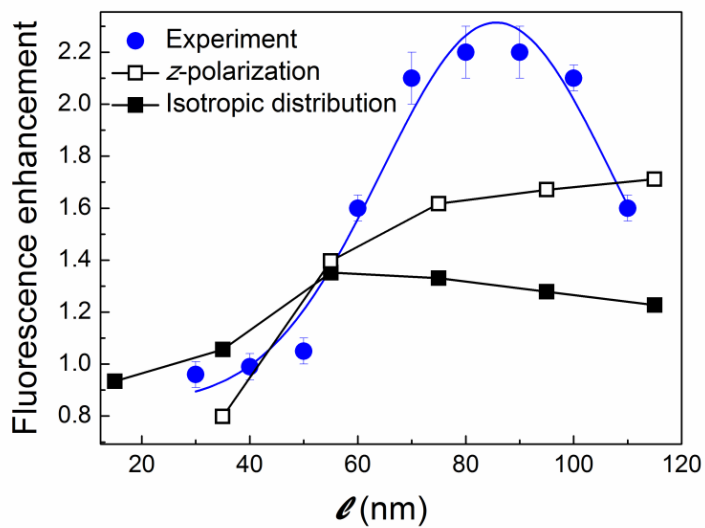

**Figure 6.** Fluorescence enhancement factor as a function of the thickness of the dielectric layer (calculated for $\eta_0$ =0.5, with a collection angle of 24°). The continuous lines serve as a guide for the eyes. Experimental data (dots) are shown together with results from simulations for *z*-polarized chromophore dipoles (empty squares) and for a population of dipoles with isotropically-distributed orientations ($\alpha_x=\alpha_y=\alpha_z$=0.33 in Eq. 2, solid squares).

# SUPPORTING INFORMATION

# Metal-Enhanced Near-Infrared Fluorescence by Micropatterned Gold Nanocages

*Andrea Camposeo,[1] Luana Persano,[1] Rita Manco,[1,2] Yi Wang,[3,4] Pompilio Del Carro,[1] Chao Zhang,[5] Zhi-Yuan Li,[5,6] Dario Pisignano,[1,2] Younan Xia[3]*

[1]Istituto Nanoscienze-CNR, Euromediterranean Center for Nanomaterial Modelling and Technology (ECMT), via Arnesano I-73100, Lecce, Italy

[2]Dipartimento di Matematica e Fisica "Ennio De Giorgi", Università del Salento, via Arnesano I-73100 Lecce, Italy

[3]The Wallace H. Coulter Department of Biomedical Engineering, Georgia Institute of Technology and Emory University, Atlanta, Georgia, 30332, United States

[4]Key Laboratory of Green Synthesis and Applications, College of Chemistry, Chongqing Normal University, Chongqing 401331, P. R. China

[5]Laboratory of Optical Physics, Institute of Physics, Chinese Academy of Sciences, Beijing 100190, P. R. China

[6]College of Physics and Optoelectronics, South China University of Technology, Guangzhou 510641, P. R. China

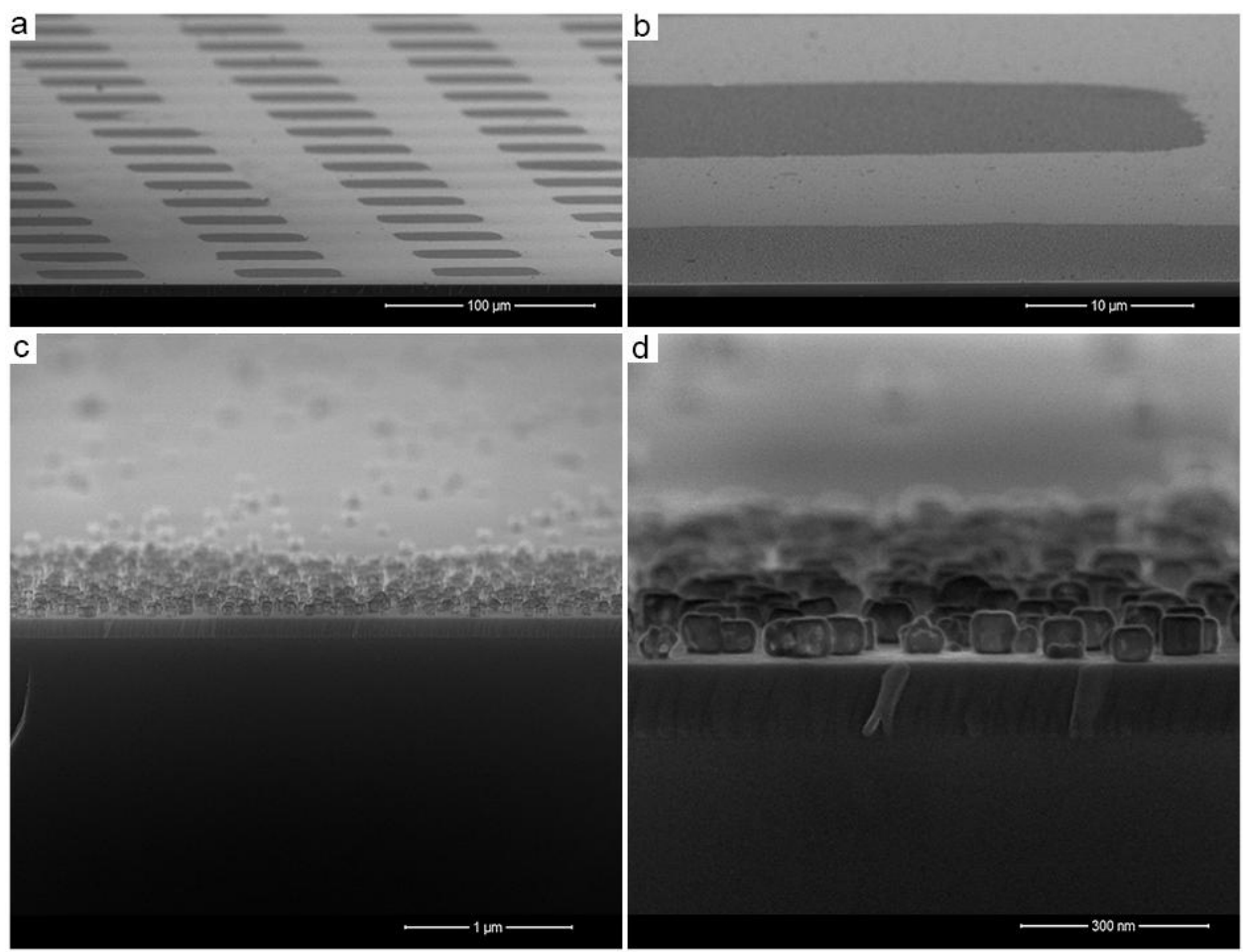

**Figure S1.** Scanning electron micrographs (at different magnifications) of a patterned array of squares whose surfaces are covered by Au nanocages with an edge length of 70 nm. The edge length of the squared islands is 50 µm.

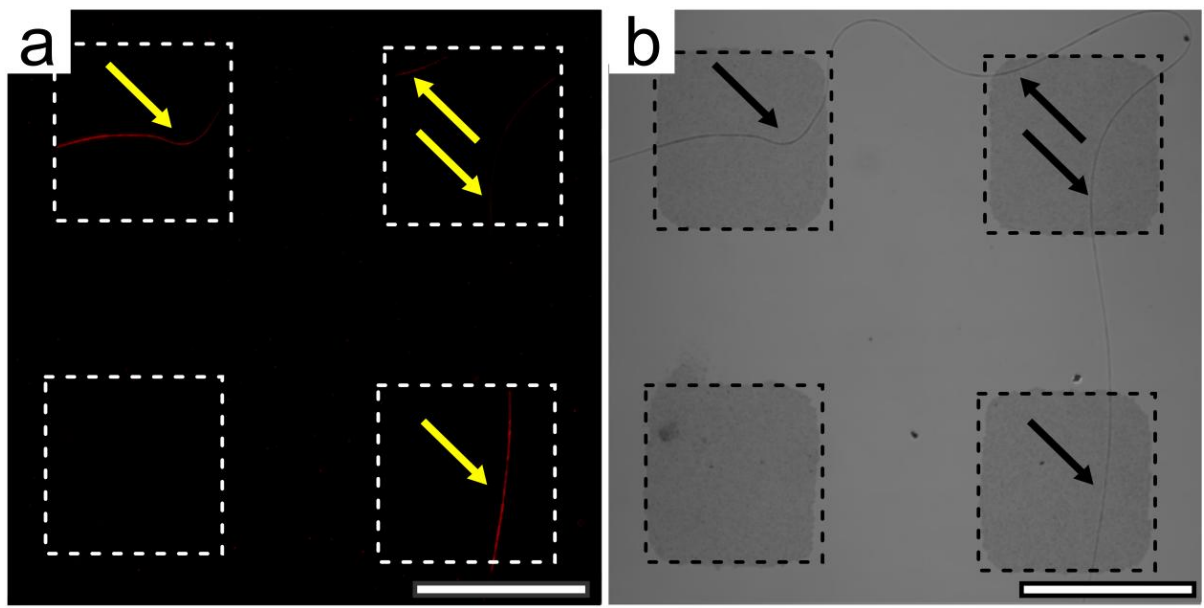

**Figure S2.** (a) Confocal fluorescence micrograph and (b) the corresponding excitation laser intensity transmission micrograph of LD 700-doped nanofibers on an array of microcontact printed squares of Au nanocages. $SiO_2$ thickness = 70 nm. Scale bars = 50 μm. Dashed lines are used to highlight the regions with patterned nanocages. The arrows indicate the MEF-enhanced segments of the light-emitting nanofibers. The emitted light is enhanced by a factor of about 4. This effect makes segments of nanofibers over the regions with Au nanocages immediately visible, with almost no background signals.

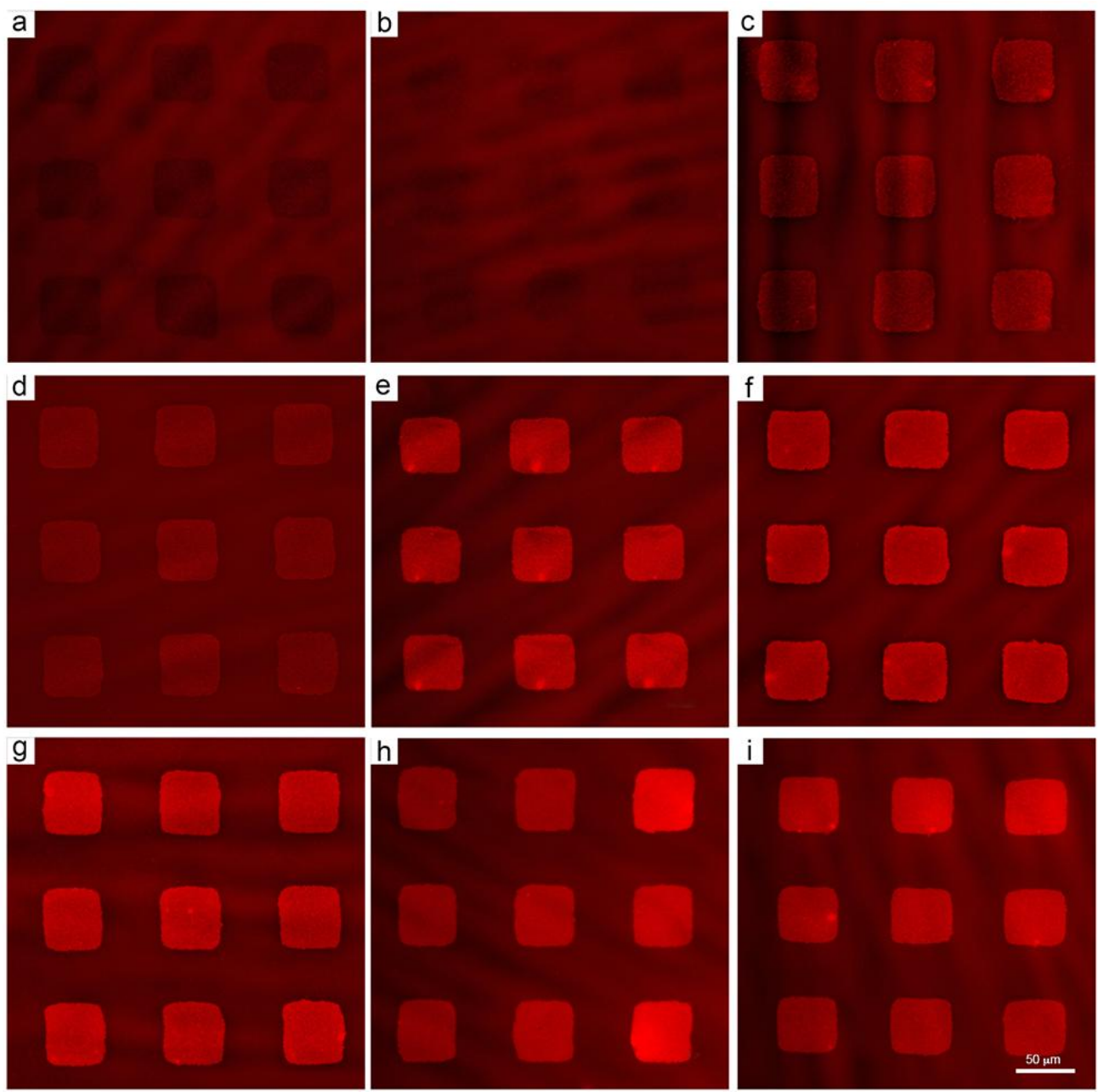

**Figure S3.** Confocal fluorescence micrographs of quartz substrates patterned with Au nanocages (on the squared islands), followed by deposition of a spacing layer of SiO$_2$ and a top layer of PMMA containing LD 700 molecules. The SiO$_2$ layers have different thickness: (a) 30 nm, (b) 40 nm, (c) 50 nm, (d) 60 nm, (e) 70 nm, (f) 80 nm, (g) 90 nm, (h) 100 nm, and (i) 110 nm.

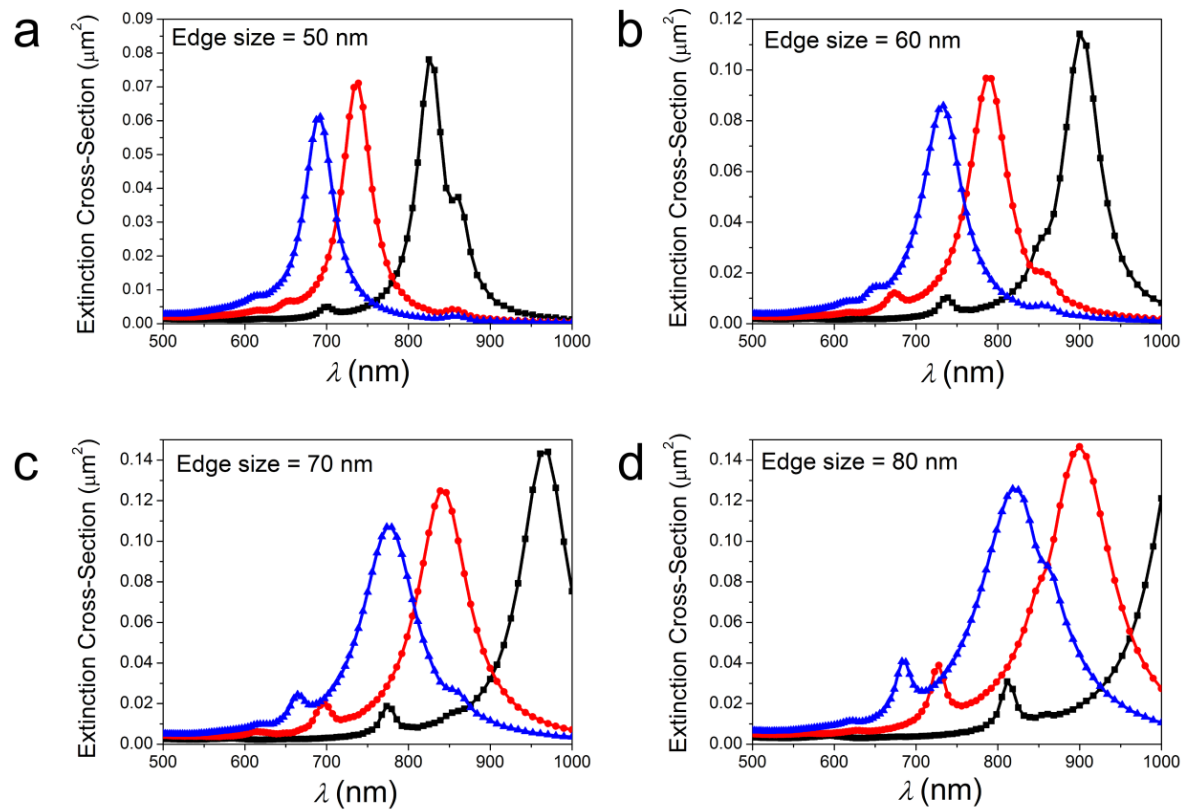

**Figure S4.** Extinction spectra calculated for nanocages with edge lengths of (a) 50 nm, (b) 60 nm, (c) 70 nm and (d) 80 nm. The wall thicknesses are 4 nm (squares, black curves), 6 nm (circles, red curves), and 8 nm (upward triangles, blue curves), respectively.

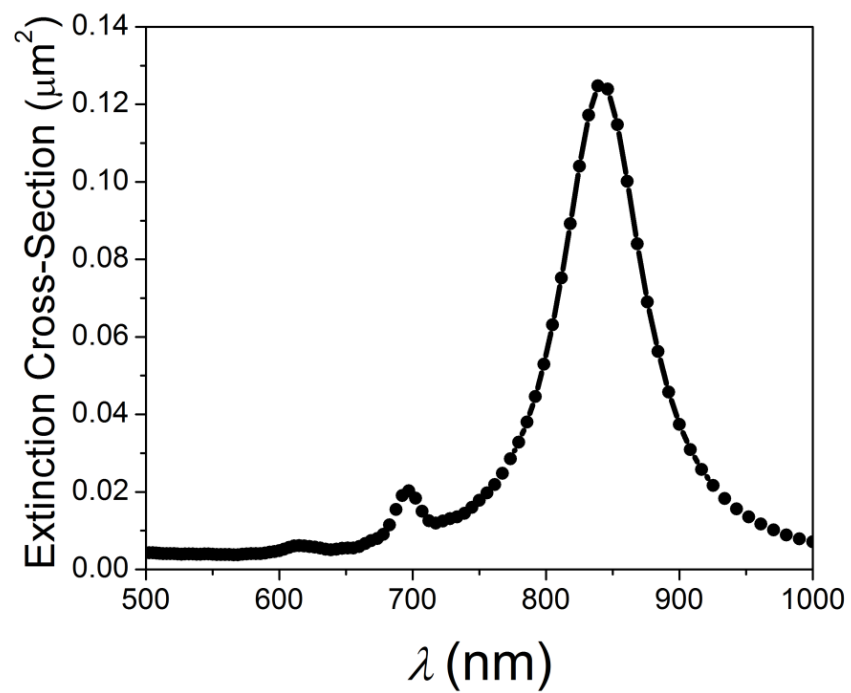

**Figure S5**. Extinction spectrum calculated for the nanocages used in this work. Edge length = 70 nm, wall thickness = 6 nm. The LSPR peak is located at about 850 nm.

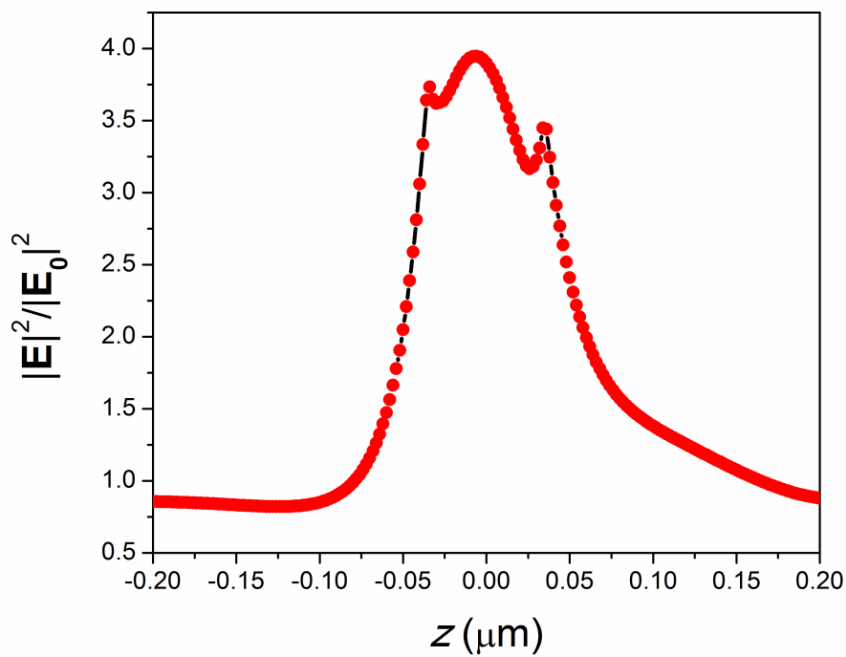

**Figure S6**. Enhancement factor for the local field intensity, calculated at different vertical positions relative to the center of the nanocage. This quantity is proportional to the excitation rate enhancement factor of the dye molecule. Excitation wavelength: 640 nm.

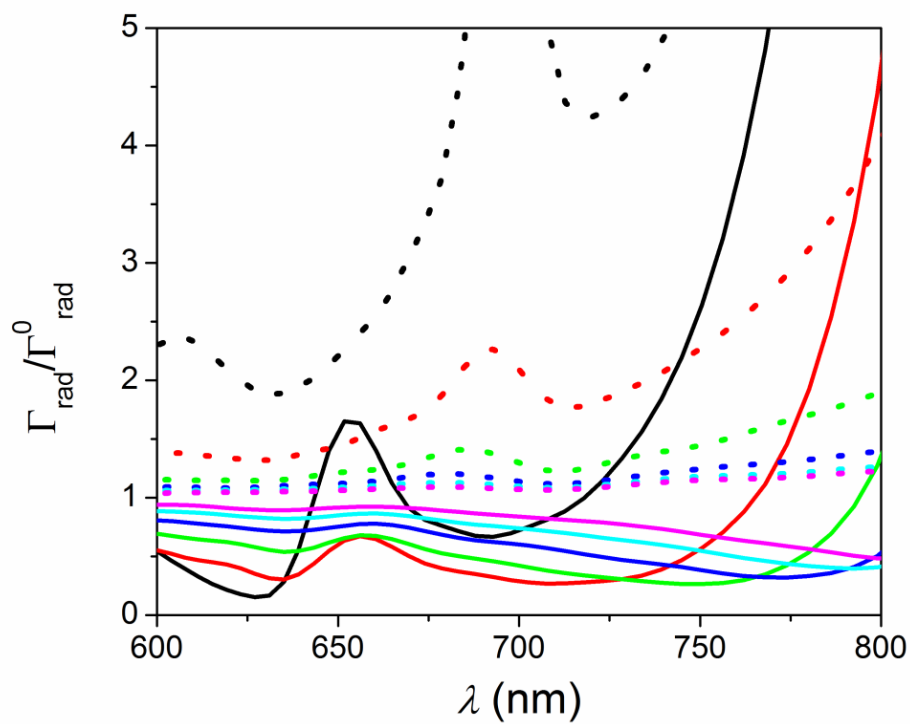

**Figure S7**. Radiation rate modified factors calculated for molecules with different dipole orientations and at different vertical distances, $l$, from the nanocage. Continuous lines: $z$-polarized dipole. Dotted lines: $x$-polarized dipole. Black, red, green, blue, cyan and magenta spectra: $l$ = 15, 35, 55, 75, 95 and 115 nm.

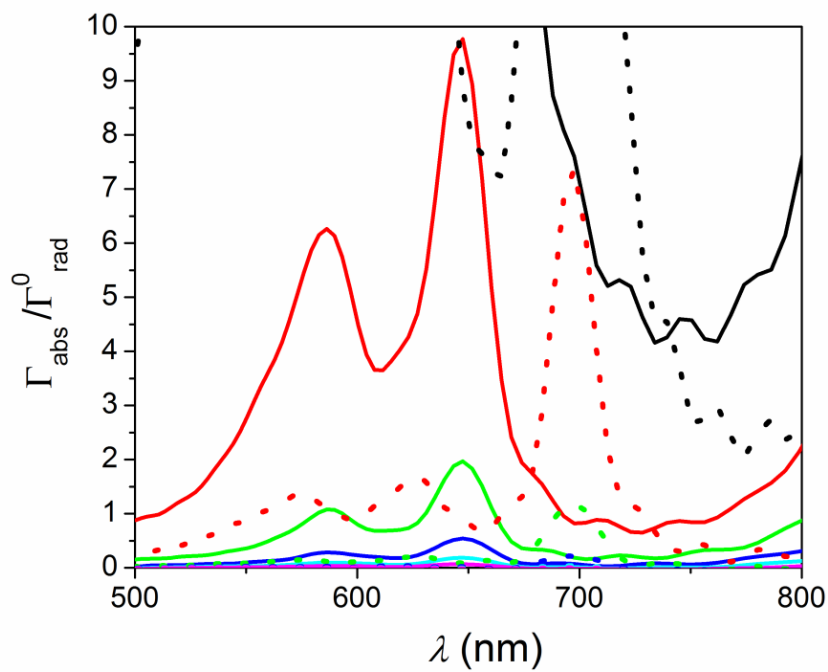

**Figure S8.** Absorption rate modified factors calculated for molecules with different dipole orientations and at different vertical distances, *l*, from the nanocage. Continuous lines: *z*-polarized dipole. Dotted lines: *x*-polarized dipole. Black, red, green, blue, cyan and magenta spectra: *l* = 15, 35, 55, 75, 95, and 115 nm.

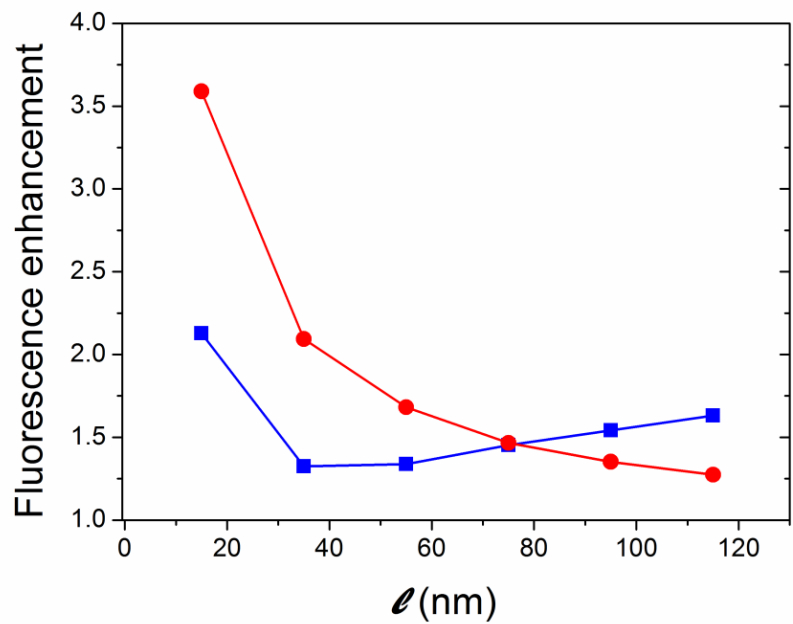

**Figure S9.** Calculated MEF enhancement factor for various nanocage-emitter distances (*l*). Here simulations are performed by considering a poorly emitting dye ( $\eta_0 \cong 0$ ). Red curve: *x*-polarized dipole. Blue curve: *z*-polarized dipole.

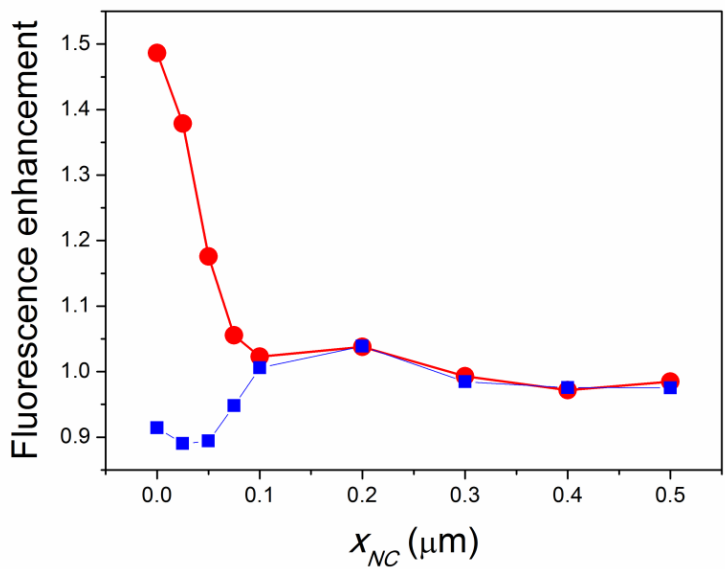

**Figure S10.** Calculated total MEF enhancement factor for various nanocage-emitter distances along the horizontal direction (here $l$ is fixed at 70 nm). $\eta_0 = 0.2$. Red curve: $x$-polarized dipole. Blue curve: $z$-polarized dipole.

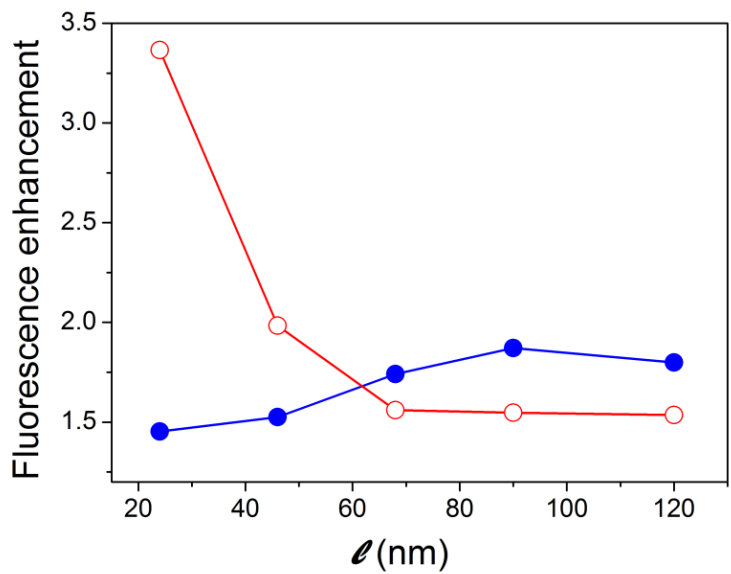

**Figure S11.** Fluorescence enhancement *vs.* thickness of the dielectric layer separating the fluorophore from the metal surface (calculated for $\eta_0$ =0.5 and by taking into account the collection angle = 24°). The lines serve as guide for the eyes. The calculated data are shown for a population of dipoles with isotropically-distributed orientations ($\alpha_x=\alpha_y=\alpha_z$=0.33 in Eq. 2), and for nanocages with a wall thickness of 8 nm and an edge length of 50 nm (open circles, red curve) and 60 nm (filled circles, blue curve), respectively.

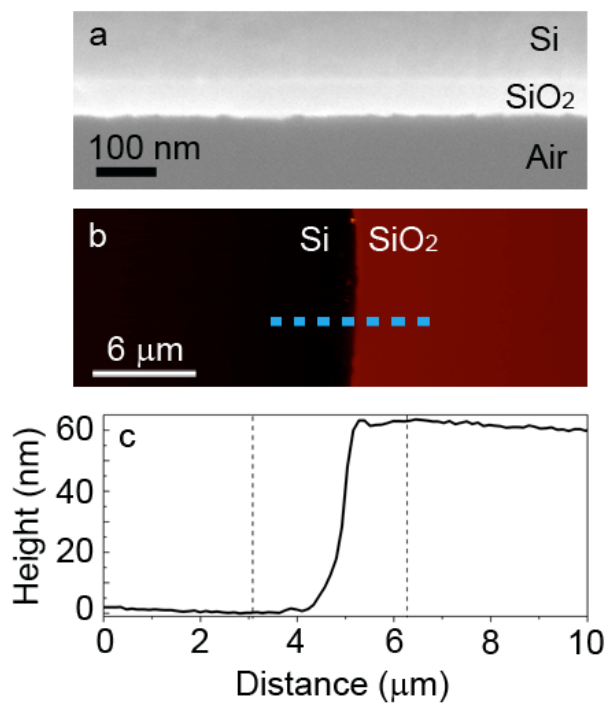

**Figure S12**. Calibration of the thickness of deposited $SiO_2$. (a) Cross-sectional SEM micrograph of a layer of $SiO_2$ deposited on Si. The measured thickness of the $SiO_2$ layer is 60 nm. (b, c) The planar and cross-sectional views of the surface by atomic force microscopy. The height profile is taken along the blue dashed line marked in (b), highlighting the Si/$SiO_2$ step between the vertical dashed lines in (c).